\documentclass[11pt]{article}

\usepackage[margin=1in]{geometry}
\usepackage{graphicx}
\usepackage{multirow}
\usepackage{amsmath,amssymb,amsfonts}
\usepackage{booktabs}
\usepackage{xcolor}
\usepackage{hyperref}
\usepackage{tabularx}
\usepackage{listings}
\usepackage{pdflscape}
\usepackage{float}
\usepackage{caption}
\usepackage{authblk}
\usepackage[numbers,sort&compress]{natbib}

\begin{document}

\title{UNISEP: A Unified Sensor Placement Framework for Human Motion Capture and Wearables}

\author[1]{Julius Welzel}
\author[2,3]{Sein Jeung}
\author[1]{Lara Godbersen}
\author[4]{Seyed Yahya Shirazi\thanks{Corresponding author: shirazi@ieee.org}}


\affil[1]{Department of Neurology, Kiel University, Kiel, Germany}
\affil[2]{Technical University of Berlin, Berlin, Germany}
\affil[3]{Max Planck Institute for Human Cognitive and Brain Sciences, Leipzig, Germany}
\affil[4]{Swartz Center for Computational Neuroscience, Institute for Neural Computation, University of California San Diego, La Jolla, CA, USA}

\renewcommand\Authfont{\normalsize}
\renewcommand\Affilfont{\small\itshape}
\date{}

\maketitle

\begin{abstract}The proliferation of wearable sensors and monitoring technologies has created a need for standardized sensor placement protocols. While existing standards like the Surface Electromyography for Non-Invasive Assessment of Muscles (SENIAM) recommendations for electromyography (EMG) and the 10--20 system for electroencephalography (EEG) address modality-specific applications, no comprehensive framework spans different sensing modalities and applications. We present the Unified Sensor Placement (UNISEP) framework to facilitate reproducible handling of human movement and physiological data across various systems and research domains. The framework provides a method to describe coordinate systems and placement protocols based on anatomical landmarks, and is designed to complement existing data-sharing standards such as the Brain Imaging Data Structure (BIDS) and Hierarchical Event Descriptors (HED). Even during its proposal stage, the UNISEP approach has been adopted by the EMG-BIDS extension (BIDS version~1.11.0), confirming the community need for a unified, machine-readable sensor placement framework. The UNISEP framework facilitates consistency, reproducibility, and interoperability in applications ranging from lab-based clinical biomechanics to continuous health monitoring in everyday life.
\end{abstract}


\section{Introduction}\label{sec:intro}

Measurement of human movement and physiological signals is fundamental in fields ranging from biomechanics to psychology. It is also instrumental in clinical rehabilitation and continuous health monitoring ~\cite{cappozzo1995position,kuderle2022placement}. Technological advancements in motion capture systems, inertial sensors, and wearable devices have expanded data acquisition and analysis opportunities that reach far beyond traditional laboratory applications. Since the placement of body-worn sensors has a significant effect on acquired signals, a standardization routine for sensor placement practices is essential. A standardized protocol is of utmost importance to improve data consistency, interpretability, reproducibility, and interoperability across diverse applications~\cite{kuderle2022placement,zhou2025imucoco,campanini2007effect,werling2023addbiomechanics,niswander2021evaluating}.

The impact of sensor placement on data quality is well-documented. In the field of biomechanics, studies have shown that even small changes in sensor positioning can significantly affect measurements and their outcomes. In electromyography (EMG), electrode placement variations relative to the target muscle can lead to signal amplitude changes of up to 50\% and altered muscle activation patterns~\cite{wong2006surface,rainoldi2004geometrical}. Similar effects have been observed with inertial measurement units (IMUs), where variations in placement and orientation can substantially impact acceleration and angular velocity measurements during dynamic movements~\cite{tan2019influence}. The same problem applies to optical motion capture applications; the deviation of markers from predefined positions can affect downstream kinematic estimates~\cite{osis2016effects}.

Accuracy of sensor placement affects data quality across body-worn measurement modalities. In the measurement of biopotentials, such as electrocardiography (ECG), EMG, and electroencephalography (EEG), electrode positioning directly affects waveform morphology and measurement accuracy, with small deviations potentially leading to misinterpretation~\cite{bond2012effects,bond2011simulation,roy2020comparison}. The same principle applies to IMU-based systems, continuous monitoring devices, and emerging wearable technologies where sensor location relative to the underlying anatomy determines data quality.

Several modality-specific standards address sensor placement within their respective domains. The Surface Electromyography for Non-Invasive Assessment of Muscles (SENIAM) recommendations provide guidelines for surface EMG electrode placement on specific muscles, including recommended positions, electrode sizes, and inter-electrode distances~\cite{hermens1999european}. The Mason-Likar configuration specifies electrode positions for exercise ECG monitoring~\cite{mason1966new,francis2016ecg}. In EEG, the international 10--20 system~\cite{klem1999ten} and the extended 5--10 system~\cite{oostenveld2001five} provide standardized electrode placement to ensure reproducibility and anatomical relevance. Yet, these standards are modality-specific and typically require expertise for correct application. This highlights a critical gap: while existing standards address specific biosignal modalities, a more comprehensive and application-agnostic framework suited for including various sensors does not exist. Moreover, current standards leave room for improvement~\cite{campanini2020surface,manca2020survey}, to incorporate a wider user base with varying levels of expertise in anatomy, biomechanics, and biosensors.

At the same time, large-scale data-sharing initiatives and machine-learning techniques in health monitoring and movement analysis require a comprehensive sensor placement framework. These applications require consistent and well-documented data collection protocols for meaningful comparisons and reliable results. Current data-sharing standards like the Brain Imaging Data Structure (BIDS)~\cite{gorgolewski2016brain} and the Hierarchical Event Descriptors (HED)~\cite{robbins2021capturing} provide robust frameworks for data organization but do not include guidelines for sensor placement across multiple types of human monitoring applications. Critically, such data standards require machine-readable metadata describing sensor positions, as conveying this information through visual documentation alone is impractical for an automated workflow.

Here, we propose the Unified Sensor Placement (UNISEP) framework addressing these challenges through a comprehensive system of anatomical landmarks, coordinate systems, and placement protocols. UNISEP defines anatomical coordinate systems for body segments based on established landmarks~\cite{cappozzo1995position,hanavan1966personalized} and establishes a hierarchical system for different measurement contexts. UNISEP is technology-agnostic, accommodating various sensing modalities while maintaining compatibility with existing data-sharing standards, allowing existing records to be annotated with the proposed framework (see examples of motion capture and IMU sensor placement \href{https://unisep-framework.github.io/anatomical_table.html}{online}).

The following sections detail the proposed framework, beginning with fundamental definitions and proceeding to specific placement protocols for different body segments with recommendations for implementation and validation.

\section{The UNISEP Framework}\label{sec:framework}

\subsection{Anatomical Coordinate Systems}\label{sec:coordsys}

The foundation of the UNISEP framework rests on the use of established anatomical landmarks and spatial references to describe Cartesian coordinate systems. These definitions enable consistent interpretation and implementation across different applications and laboratories.

A coordinate system consists of an origin point and a set of axes that define directions in space. In human movement analysis, we encounter multiple coordinate systems: the global laboratory system, anatomical systems tied to body segments, and sensor- or device-specific systems. The relationships between these systems must be clearly defined to ensure meaningful data interpretation. These practices are consistent with the International Society of Biomechanics (ISB) recommendations for standardization of kinematic data reporting, which emphasize clear definitions of anatomical and global coordinate systems~\cite{wu1995isb,wu2002isb}.

The anatomical coordinate system for each body segment is defined using palpable landmarks that can be reliably identified across individuals. These landmarks, drawn from established anatomical references (see Section~\ref{sec:landmarks} and Appendix~A), are used to define the coordinate system of a body segment and to provide spatial reference for sensor placement. The framework defines each coordinate system through:

\begin{enumerate}
    \item An origin point based on specific anatomical landmarks.
    \item Primary axes aligned with functional anatomical directions.
    \item Clear definitions of positive directions and measurement dimensions (e.g., percent of segment length, millimeters).
\end{enumerate}

The framework defines sensor placement locations using local anatomical coordinate systems. These local systems are part of a larger kinematic chain, consistent with practices in biomechanics and robotics where transformations between local and global coordinate systems are computed through forward kinematics~\cite{cappozzo1995position}. Although the framework focuses on local definitions, users can derive global coordinates through standard transformation techniques if needed.

\subsection{Anatomical Landmark System}\label{sec:landmarks}

The framework uses a set of anatomical landmarks drawn from established anatomical and biomechanical references~\cite{cappozzo1995position,wu1995isb,wu2002isb}. The landmarks were selected for their reliability, accessibility across different body types, relevance to common sensor placement needs, and minimal displacement during movement. Each landmark is defined using standardized anatomical terminology and palpation methods. The complete set of landmarks and their definitions is provided in the anatomical landmark table (\href{https://unisep-framework.github.io/anatomical_table.html}{online}, and Appendix~A), forming the backbone of the framework.

\subsection{Sensor Placement Protocol}\label{sec:protocol}

The core of the framework is a unified placement protocol that defines sensor locations relative to anatomical landmarks using standardized coordinate systems. Each sensor location is specified in three steps:

\begin{enumerate}
    \item Identify the relevant body segment and its anatomical coordinate system.
    \item Determine the location using normalized coordinates within that local body segment coordinate system (0--100\% along each axis), enabling the placement to scale with the individual's body proportions.
    \item Specify the measurement method (visual inspection, measuring tape, 3D scanning, etc.) and the standard used during placement (SENIAM, 10--20, etc.).
\end{enumerate}

This approach differs from existing modality-specific standards in two key ways. First, the normalized coordinate system (0--100\% along each axis) scales placement descriptions to different body sizes and proportions, enabling transferability across subjects. Second, the framework provides a common spatial language across sensing modalities: an EMG electrode, an IMU, and a motion capture marker placed on the same body segment can all be described within the same coordinate system, facilitating documentation of multi-sensor setups.

The protocol provides a foundation for consistent sensor placement while maintaining the flexibility needed for diverse applications in human movement analysis.

\section{Anatomical Landmark Definitions}\label{sec:landmark_defs}

The practical implementation of the framework relies on anatomical landmarks and their relationships. While the complete system covers all major body segments, we present here the thorax/upper torso as an exemplar of the approach (Table~\ref{tab:thorax}).

As a worked example, consider placing an ECG electrode on the chest. Within the thorax coordinate system defined by the Left Acromion Process (LAP), Right Acromion Process (RAP), C7 vertebra, and Xiphoid Process, a V1 electrode position (fourth intercostal space, right sternal border) could be described as approximately $X = 40\%$, $Y = 0\%$ (anterior surface), $Z = 60\%$ from the C7 origin. This encoding is both human-readable and machine-readable, enabling automated comparison across studies and integration with data-sharing standards.

The complete table for all body segments is available at \href{https://unisep-framework.github.io/anatomical_table.html}{unisep-framework.github.io} and in the supplementary materials.

\begin{figure}[ht]
\centering
\includegraphics[width=0.35\textwidth]{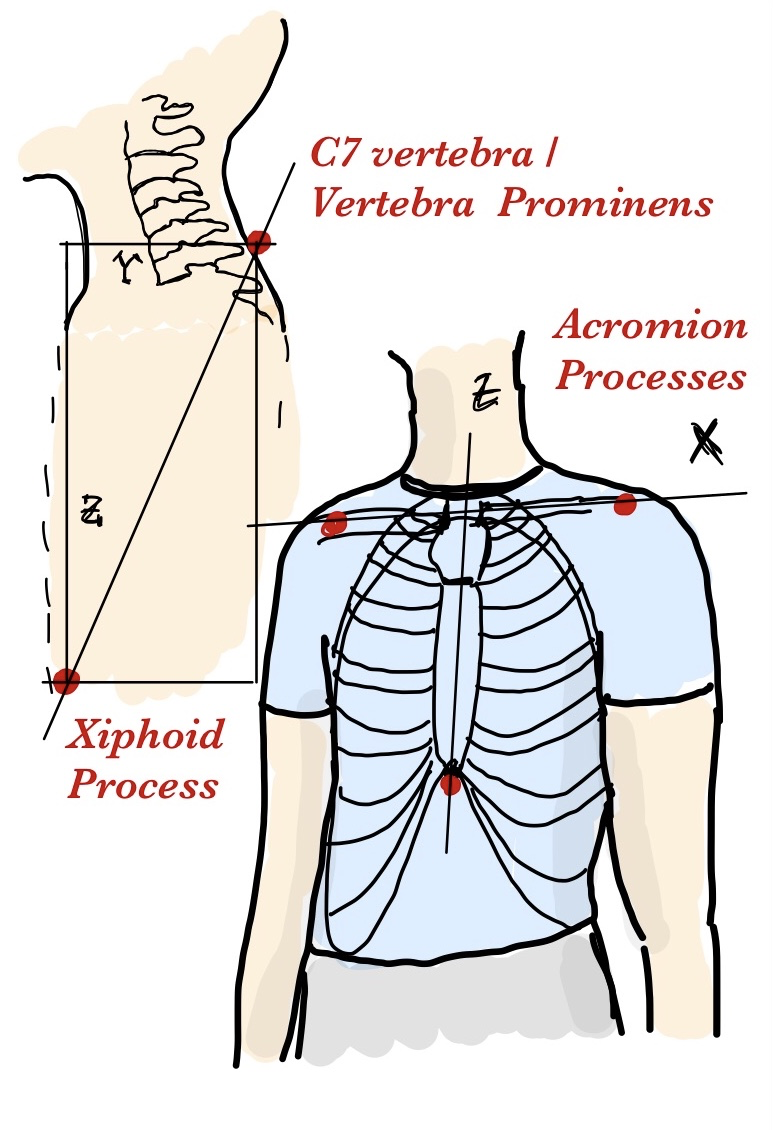}
\caption{Landmark visualization for the thorax/upper torso coordinate system. The coordinate system is defined by four palpable anatomical landmarks: Left and Right Acromion Processes (LAP, RAP), C7 vertebra, and Xiphoid Process.}\label{fig:thorax}
\end{figure}

\begin{table}[ht]
\caption{Example implementation for the thorax/upper torso. The coordinate system is defined by four palpable anatomical landmarks. Sensor positions are expressed as normalized percentages along each axis, enabling the description to scale with body size. See Figure~\ref{fig:thorax} for visualization.}\label{tab:thorax}
\begin{tabular*}{\textwidth}{@{\extracolsep\fill}p{2cm}p{3.5cm}p{2.5cm}p{3.5cm}}
\toprule
\textbf{Body Segment} & \textbf{Anatomical Description} & \textbf{Anatomical Landmarks} & \textbf{Coordinate System} \\
\midrule
Torso-chest & The upper part of the torso, extending from the base of the neck to the diaphragm, framed by the rib cage (ribs, sternum, thoracic vertebrae). &
Left Acromion Process (LAP), Right Acromion Process (RAP), C7 vertebra (C7), Xiphoid Process &
X: LAP $\to$ RAP; Y: C7 $\to$ Xiphoid Process (shorter axis); Z: C7 $\to$ Xiphoid Process (longer axis) \\
\bottomrule
\end{tabular*}
\end{table}

This example illustrates the essential aspects of the framework. The anatomical definition provides a clear description of the body segment and its boundaries using standard anatomical terminology. The selected landmarks consist of palpable anatomical points that form a stable coordinate system. The coordinate system definition specifies unambiguous axes using landmark pairs, with clear directional conventions. Sensor positions are expressed as percentages along each defined axis.

The same systematic approach is applied to all body segments in the complete reference table. Each entry maintains this structure while accounting for segment-specific anatomical considerations. The complete table includes definitions for 15~major body segments, covering the full body from head to feet.

\section{Use Case: EMG-BIDS}\label{sec:usecase}

The need for a unified sensor placement framework has been confirmed through its early adoption by EMG-BIDS, which is part of BIDS version~1.11.0~\cite{bids2026v1110} (the EMG-BIDS manuscript is in preparation). Even during the UNISEP proposal stage, the EMG-BIDS development process identified the same gap that UNISEP addresses: EMG electrodes can be placed anywhere on the body, and existing standards like SENIAM provide human-readable guidelines but no machine-readable format for encoding placement information alongside the data.

EMG-BIDS adopted the UNISEP approach of hierarchical coordinate systems to solve this problem. In multi-sensor recordings, where high-density electrode arrays, bipolar sensors, and motion capture markers may be placed simultaneously on the same subject, each sensor type requires its own spatial description. The UNISEP framework provides the common spatial language for this.

\begin{figure}[ht]
\centering
\includegraphics[height=3in]{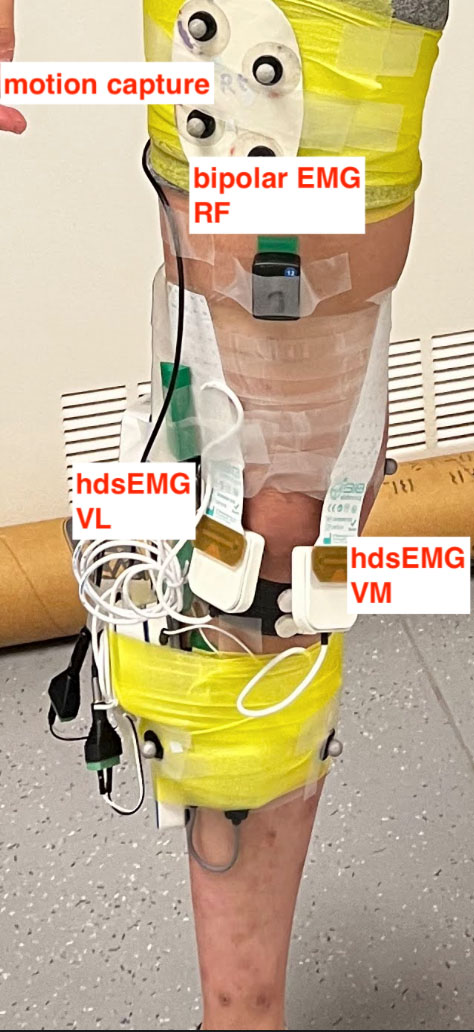}
\caption{Multi-device EMG recording setup from the EMG-BIDS specification examples. The setup includes motion capture markers, a bipolar EMG sensor on the rectus femoris (RF), and two high-density surface EMG (HD-sEMG) arrays on the vastus lateralis (VL) and vastus medialis (VM). UNISEP's hierarchical coordinate system approach enables structured documentation of all sensor positions within a unified framework. Figure adapted from the EMG-BIDS specification (in preparation).}\label{fig:emgbids_example}
\end{figure}

As a concrete example from the EMG-BIDS specification, consider a recording setup with two HD-sEMG grids on the thigh (Figure~\ref{fig:emgbids_example}). The parent coordinate system is defined using UNISEP's anatomical landmarks for the thigh segment (Femur Medial Epicondyle, Femur Lateral Epicondyle, Greater Trochanter) with normalized coordinates in percent. Each HD-sEMG grid is then described in a child coordinate system (in millimeters) that is anchored to the parent system via a specific electrode and its coordinates, described in the "Anchorcoordinates" field. This hierarchical approach enables:

\begin{itemize}
    \item Precise documentation of the physical electrode layout within each grid (in mm).
    \item Reproducible placement of the grid on the body using anatomical landmarks (in normalized \%).
    \item Machine-readable encoding in JSON sidecar files that accompany the data.
    \item Transferability across subjects with different body proportions.
\end{itemize}

The resulting metadata structure can be represented in a BIDS-compatible coordinate system file:

\begin{small}
\begin{verbatim}
{
  "EMGCoordinateSystem": "Other",
  "EMGCoordinateUnits": "mm",
  "EMGCoordinateSystemDescription":
    "Local grid coordinate system for VL array",
  "ParentCoordinateSystem": "space-thigh",
  "AnchorElectrode": "E6",
  "AnchorCoordinates": [20, 100, 50]
}
\end{verbatim}
\end{small}

\noindent where the parent \texttt{space-thigh} coordinate system is defined using UNISEP's anatomical landmarks with normalized coordinates (in percent).

This adoption demonstrates two key aspects of the UNISEP framework. First, the framework addresses a real and recognized need: the BIDS community independently arrived at the same requirement for structured, machine-readable sensor placement documentation. Second, the hierarchical coordinate system design is sufficiently flexible to accommodate both simple placements (a single bipolar sensor documented by muscle name and guideline reference) and complex configurations (multi-grid arrays with precise spatial encoding).

\section{Discussion}\label{sec:discussion}

We present a framework for standardizing sensor placement documentation in human movement and physiological monitoring applications. Through definitions of anatomical landmarks, coordinate systems, and placement protocols, the framework enables reproducible sensor placement across different applications and laboratories. The approach offers a systematic method for documenting and reproducing sensor placements through normalized coordinate systems.

\subsection{Relationship to Existing Standards}

The UNISEP framework complements, rather than replaces, existing modality-specific standards. SENIAM for EMG~\cite{hermens1999european}, Mason-Likar for ECG~\cite{mason1966new}, and the 10--20 system for EEG~\cite{klem1999ten} serve their specific applications well and have established communities of practice. The UNISEP framework provides a common spatial language that can encode placements from any of these standards within a unified coordinate system. This enables documentation of multimodal setups, where, for example, EMG electrodes following SENIAM guidelines and IMUs are placed on the same body segment and must be described in relation to each other.

This approach aligns with how the ISB has standardized kinematic reporting through anatomical coordinate systems~\cite{wu1995isb,wu2002isb}, and extends a similar philosophy to the broader domain of sensor placement.

\subsection{Machine-Readability and Data Sharing}

A key motivation for the UNISEP framework is the need for machine-readable sensor placement metadata in modern data-sharing ecosystems. Visual documentation (photographs, diagrams) remains valuable for human understanding, but cannot be parsed by automated pipelines. Data standards such as BIDS~\cite{gorgolewski2016brain} and HED~\cite{robbins2021capturing} require structured metadata in standardized formats (JSON, TSV) to enable programmatic access, validation, and cross-study comparison.

The UNISEP framework provides the common spatial language needed for such structured encoding. Its early adoption by EMG-BIDS and conceptual alignment with Motion-BIDS~\cite{jeung2024motion} suggest that a unified coordinate system approach is a natural fit for the BIDS ecosystem. This standardization enhances data FAIRness (Findability, Accessibility, Interoperability, and Reusability)~\cite{wilkinson2016fair} by enabling clear documentation and facilitating data sharing across research groups and applications.

\subsection{Sensor Orientation}

An important aspect not yet addressed in this framework is sensor orientation. While the placement protocol provides anatomical landmarks and spatial coordinates for defining sensor locations, many sensing modalities, particularly IMUs, are highly sensitive to orientation relative to anatomical axes. Incorrect or inconsistent orientation can introduce systematic biases, distort signal interpretation, or complicate cross-study comparisons, even when placement location is consistent. While there are approaches to address this limitation~\cite{cereatti2024isb}, future iterations of the framework should include explicit orientation conventions, such as aligning sensor axes with anatomical coordinate systems or using standardized calibration procedures. This addition would extend the reproducibility of the framework beyond placement location and enable more robust integration of multimodal datasets.

\subsection{Limitations and Future Work}

Limitations of this framework include operator dependency and anatomical variability. While the framework provides clear definitions and measurement protocols, achieving consistent placement depends on operator expertise and training. The framework cannot fully address the subjective aspects of anatomical landmark identification, particularly in subjects with varying body compositions. Additionally, the current focus on static placement may require adaptation for dynamic applications where sensor position might change during movement. Recent initiatives have highlighted the importance of rigorous validation and consensus in defining biomechanical standards~\cite{cereatti2024isb}, underscoring the need to extend similar community-driven approaches to sensor placement.

Several developments are needed to enhance the practical implementation of the framework. First, formal vocabulary specifications through standard bodies such as BIDS and HED would provide cross-references and specific guidelines based on their established norms. A validation study involving multiple operators placing sensors according to the guidelines would quantify inter-operator reliability. Explicit mappings between UNISEP and existing standards like SENIAM would facilitate adoption. Furthermore, software tools for coordinate calculation and placement visualization would support practical implementation.

\subsection{Conclusion}

The UNISEP framework addresses a recognized need for unified sensor placement documentation across sensing modalities. By providing normalized coordinate systems based on anatomical landmarks, the framework enables both human-readable and machine-readable descriptions of sensor locations. Its early adoption by EMG-BIDS, even during the proposal stage, confirms the community demand for such a standard. As wearable technology continues to advance and integrate multiple sensing modalities, we anticipate that a standardized spatial language for sensor placement will become increasingly important. We provide a foundation for this standardization and welcome community feedback to evolve the framework alongside technological advancement.

\section*{Code Availability}

All information about the framework is available at \href{https://unisep-framework.github.io}{unisep-framework.github.io} and the source files are available at the corresponding GitHub repository.\footnote{\url{https://github.com/unisep-framework/unisep-framework.github.io/tree/v0.1.0-alpha}}

\section*{Author Contributions}

JW proposed the framework. SYS, JW, and SJ formalized the framework structure. SYS formulated the anatomical landmarks. SYS, JW, and SJ drafted the framework and the manuscript. LG created the graphics.

\section*{Declarations}

\begin{itemize}
\item \textbf{Competing interests}: SYS is a core contributor to the EMG-BIDS specification, which adopted the UNISEP approach during its development. The EMG-BIDS specification was developed through community consensus. The remaining authors declare no competing interests.
\item \textbf{Data availability}: The complete anatomical landmark table and placement examples are available at \url{https://unisep-framework.github.io}.
\end{itemize}

\bibliography{unisep}

\appendix

\section{The Anatomical Landmark Table}\label{app:landmarks}

The complete anatomical landmark table defines coordinate systems for 15~body segments. Each entry specifies the body segment, anatomical description, palpable landmarks (fiducials), coordinate system axes, and a landmark visualization. The full table with interactive examples for IMU and motion capture placement is available at \href{https://unisep-framework.github.io/anatomical_table.html}{unisep-framework.github.io/anatomical\_table.html}.
Tables~\ref{tab:landmarks} and~\ref{tab:landmarks2} present the complete set of body segment definitions.

\begin{landscape}
\begin{table}[H]
\caption{Anatomical landmark definitions for body segments (Part 1: Head, Neck, Torso, Abdomen, and Hand). Each segment is defined by palpable landmarks forming a local coordinate system with normalized percentages (0--100\%).}\label{tab:landmarks}
\vspace{-0.5em}
\scriptsize
\setlength{\tabcolsep}{3pt}
\begin{tabular*}{\linewidth}{@{\extracolsep\fill}clp{2.5cm}p{3.2cm}p{2.8cm}c}
\toprule
\textbf{\#} & \textbf{HED Label} & \textbf{Fiducials} & \textbf{Coordinate System} & \textbf{Description} & \textbf{Visualization} \\
\midrule
1 & Head & Nasion, Inion, LHJ, RHJ, Vertex, MMP & X: LHJ $\to$ RHJ; Y: Inion $\to$ Nasion; Z: MMP $\to$ Vertex & Superior to the neck, encasing the brain & \includegraphics[height=0.7in]{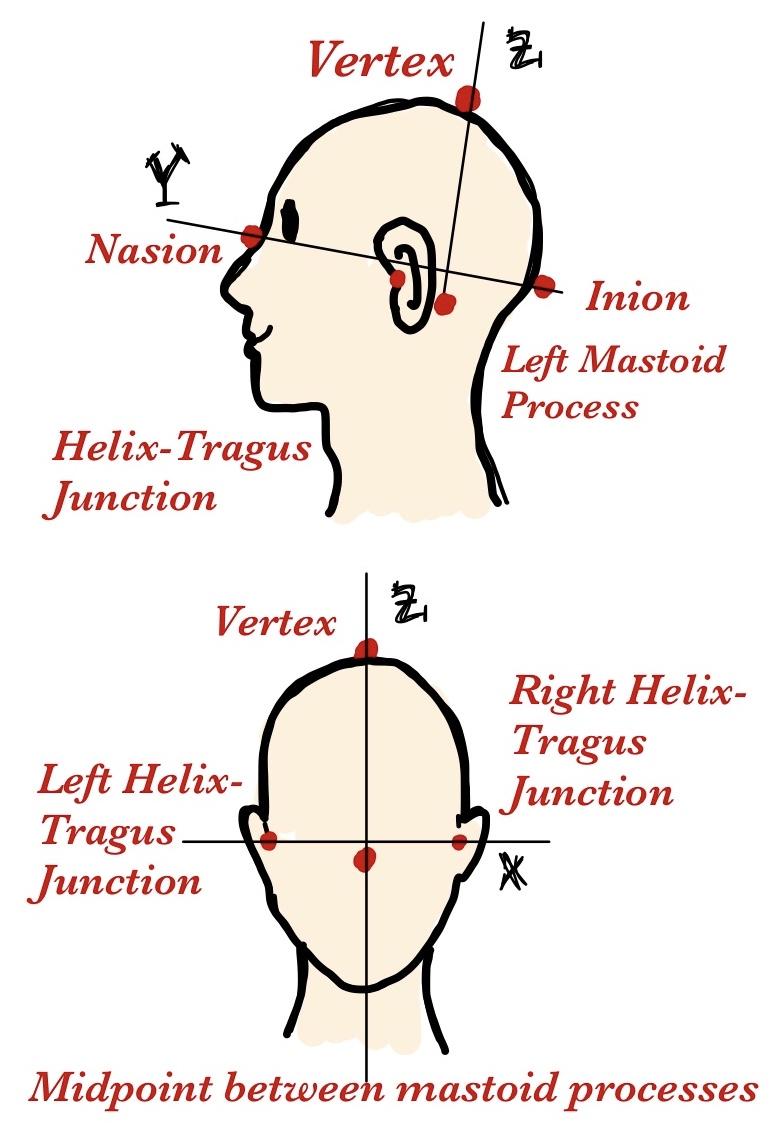} \\
\addlinespace[2pt]
n/a & Neck & C7, Jugular Notch, LMP, RMP, MMP & X: LMP $\to$ RMP; Y: C7 $\to$ Jugular Notch; Z: C7 $\to$ MMP & Bounded by base of skull and clavicles & \includegraphics[height=0.7in]{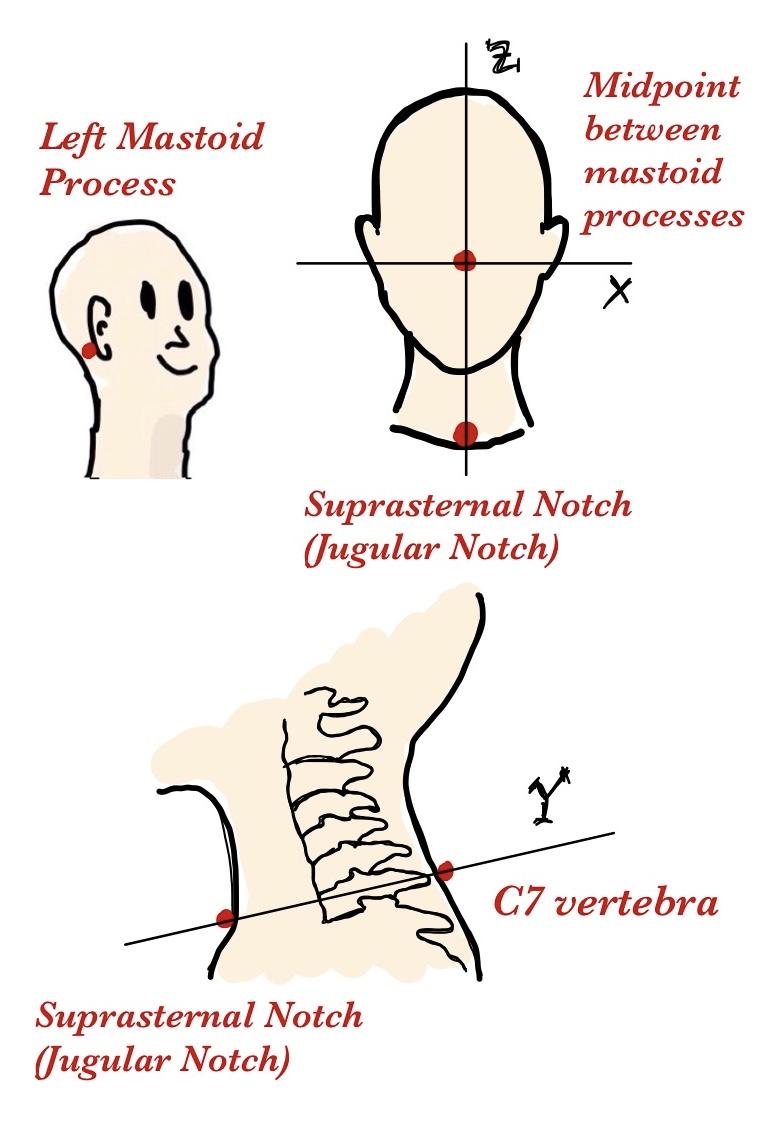} \\
\addlinespace[2pt]
2 & Torso-chest & LAP, RAP, C7, Xiphoid Process & X: LAP $\to$ RAP; Y: C7 $\to$ Xiphoid (short); Z: C7 $\to$ Xiphoid (long) & Upper torso, base of neck to diaphragm & \includegraphics[height=0.7in]{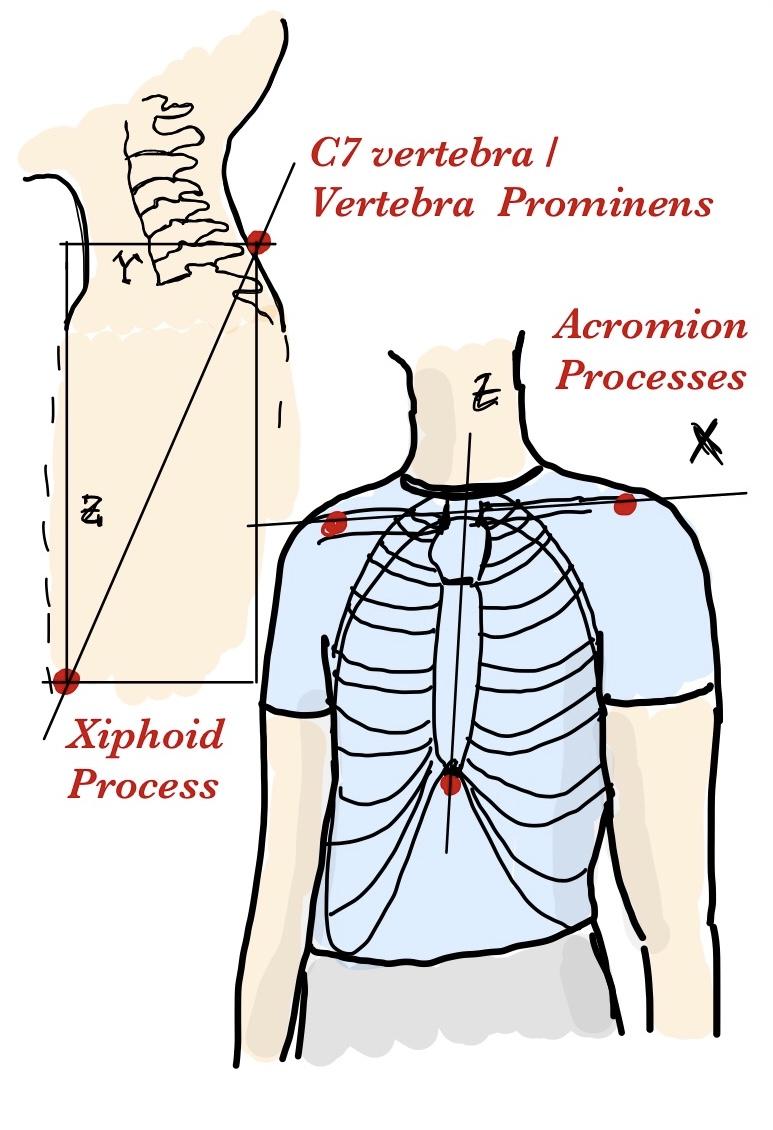} \\
\addlinespace[2pt]
3 & Abdomen & Navel, L-ASIS, R-ASIS, PSIS, Xiphoid & X: L-ASIS $\to$ R-ASIS; Y: mid-ASIS $\to$ mid-PSIS; Z: Navel $\to$ Xiphoid & Lower torso, between thorax and pelvis & \includegraphics[height=0.7in]{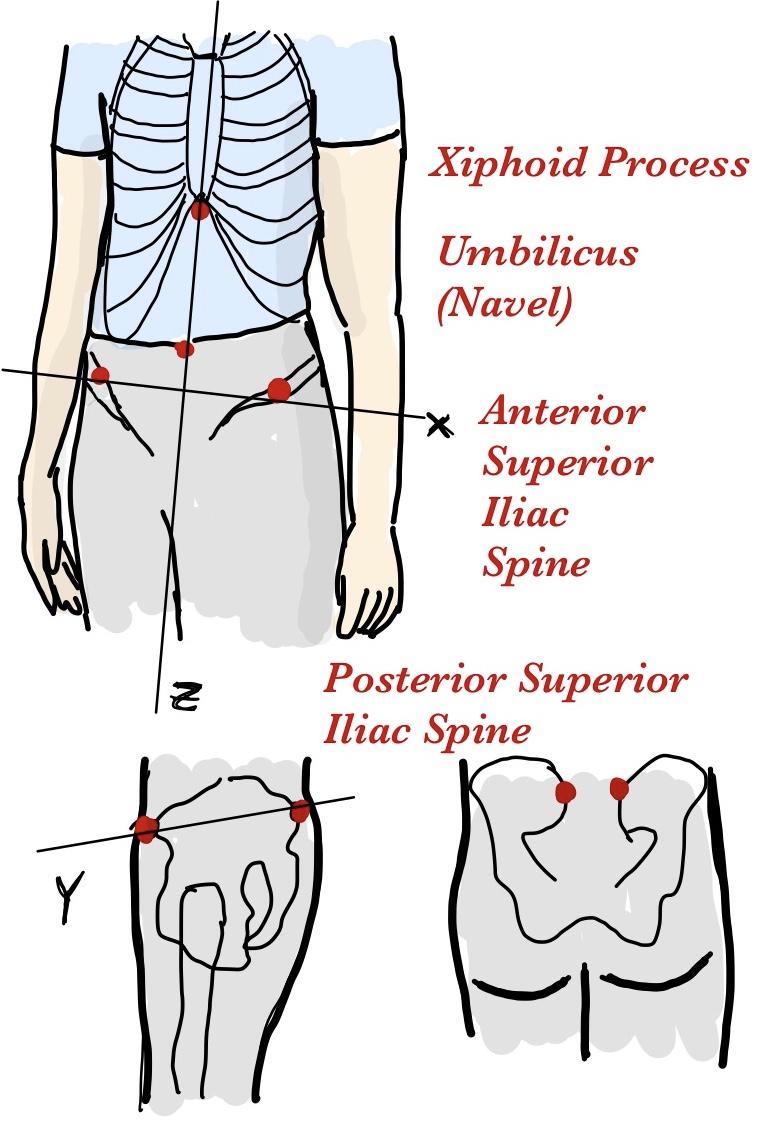} \\
\addlinespace[2pt]
4,5 & (R/L) Hand & RSP, USP, MCP & X: RSP $\leftrightarrow$ USP; Y: Dorsal $\to$ Palmar; Z: MCP $\to$ mid RSP-USP & Wrist to fingertips & \includegraphics[height=0.7in]{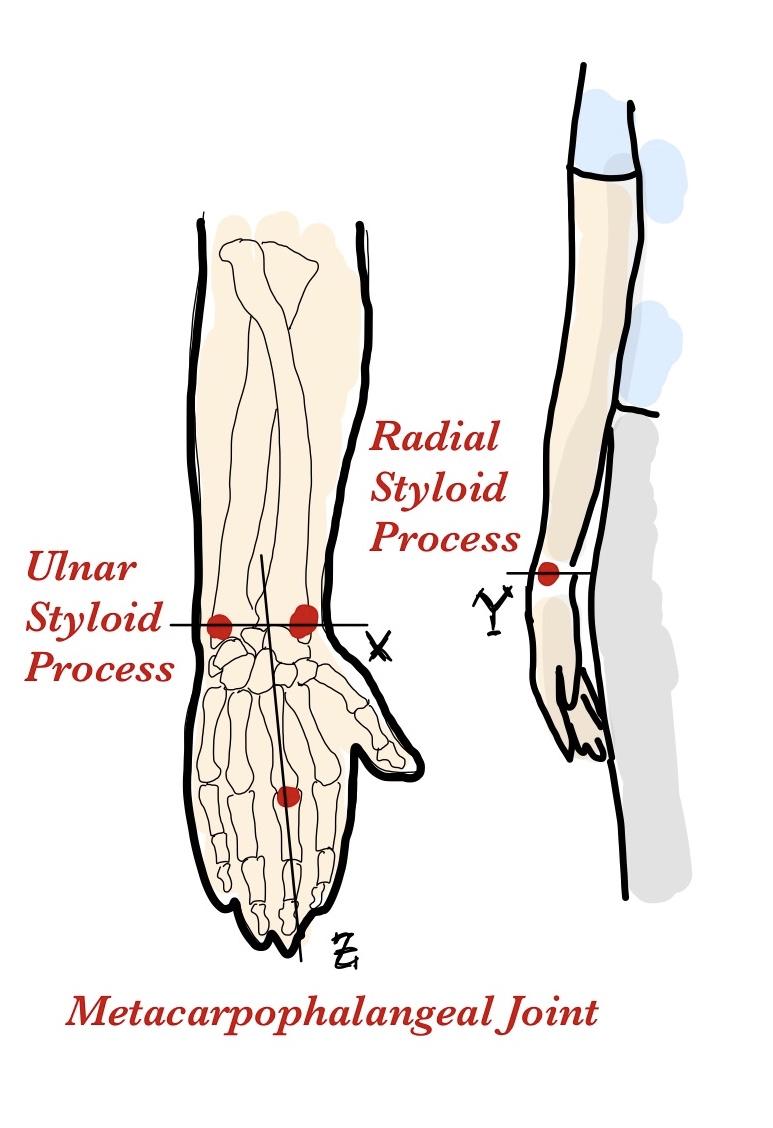} \\
\bottomrule
\end{tabular*}
\end{table}

\newpage

\begin{table}[H]
\caption{Anatomical landmark definitions for body segments (Part 2: Upper-arm, Forearm, Thigh, Lower-leg, and Foot). Continued from Table~\ref{tab:landmarks}.}\label{tab:landmarks2}
\vspace{-0.5em}
\scriptsize
\setlength{\tabcolsep}{3pt}
\begin{tabular*}{\linewidth}{@{\extracolsep\fill}clp{2.5cm}p{3.2cm}p{2.8cm}c}
\toprule
\textbf{\#} & \textbf{HED Label} & \textbf{Fiducials} & \textbf{Coordinate System} & \textbf{Description} & \textbf{Visualization} \\
\midrule
6,7 & (R/L) Upper-arm & Acromion, MHE, LHE, Olecranon, Cubital Fossa & X: MHE $\to$ LHE; Y: Olecranon $\to$ Cubital Fossa; Z: mid MHE-LHE $\to$ Acromion & Shoulder to elbow & \includegraphics[height=0.7in]{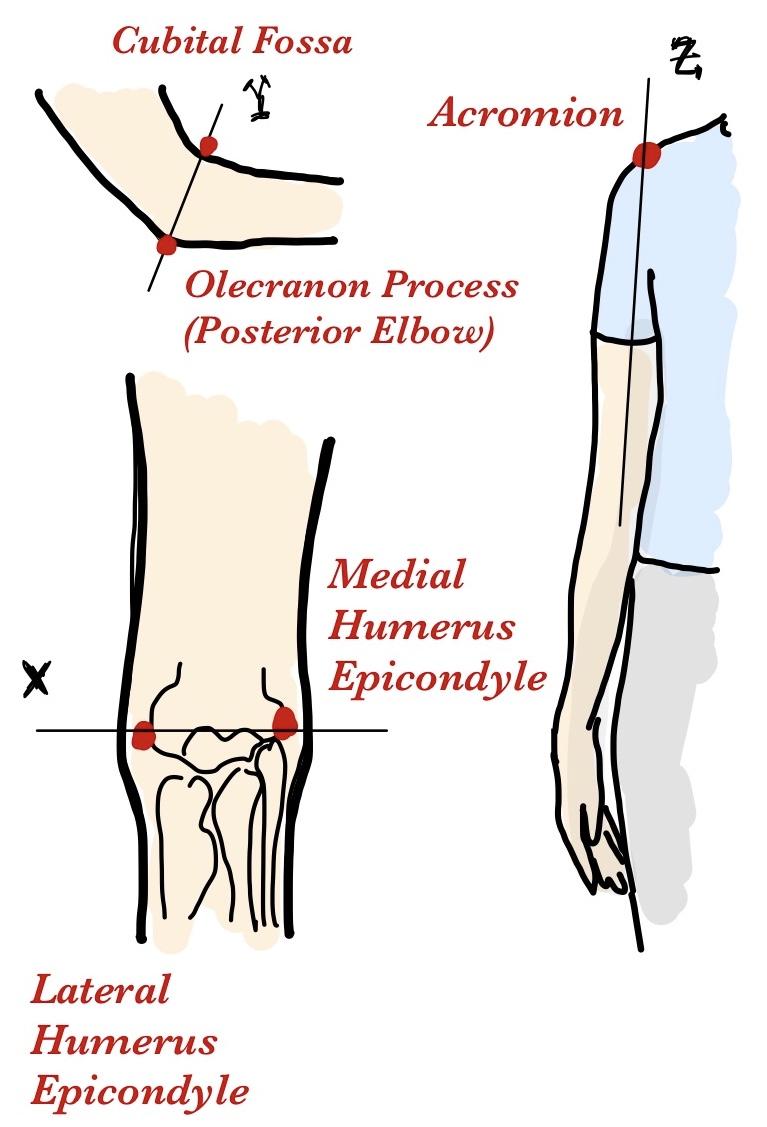} \\
\addlinespace[2pt]
8,9 & (R/L) Forearm & RSP, USP, LHE, Olecranon, Cubital Fossa & X: RSP $\leftrightarrow$ USP; Y: right-hand rule; Z: mid RSP-USP $\to$ LHE & Elbow to wrist & \includegraphics[height=0.7in]{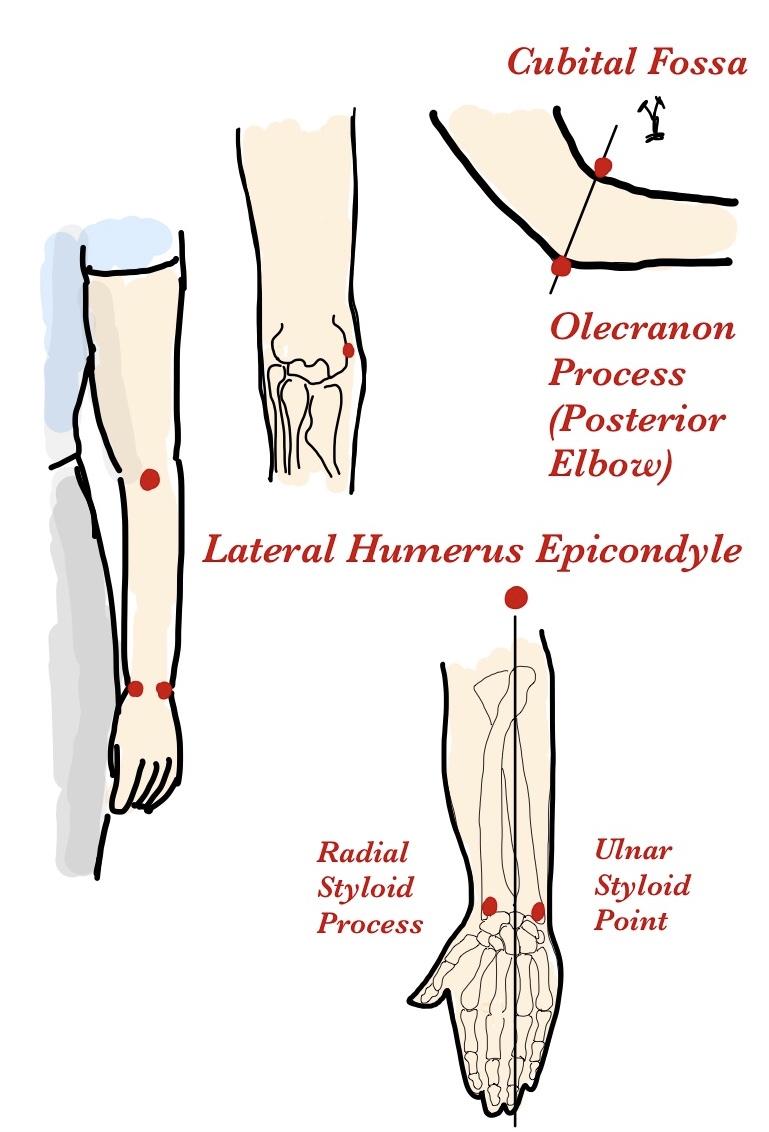} \\
\addlinespace[2pt]
10,11 & (R/L) Thigh & Greater Trochanter, FME, FLE, Patellar Surface, Popliteal Fossa & X: FME $\leftrightarrow$ FLE; Y: Popliteal $\to$ Patellar; Z: mid FME-FLE $\to$ Greater Trochanter & Hip to knee & \includegraphics[height=0.7in]{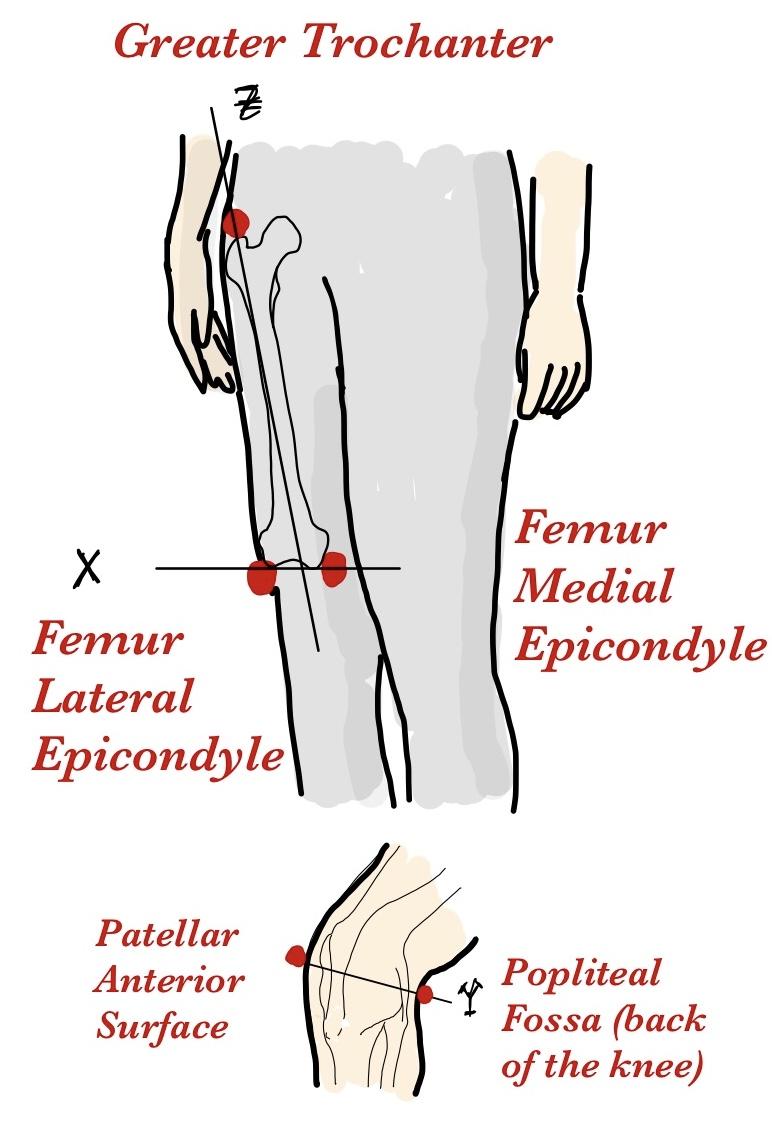} \\
\addlinespace[2pt]
12,13 & (R/L) Lower-leg & LM, MM, Patellar Surface, Popliteal Fossa & X: MM $\leftrightarrow$ LM; Y: Popliteal $\to$ Patellar; Z: mid MM-LM $\to$ Patellar & Knee to ankle & \includegraphics[height=0.7in]{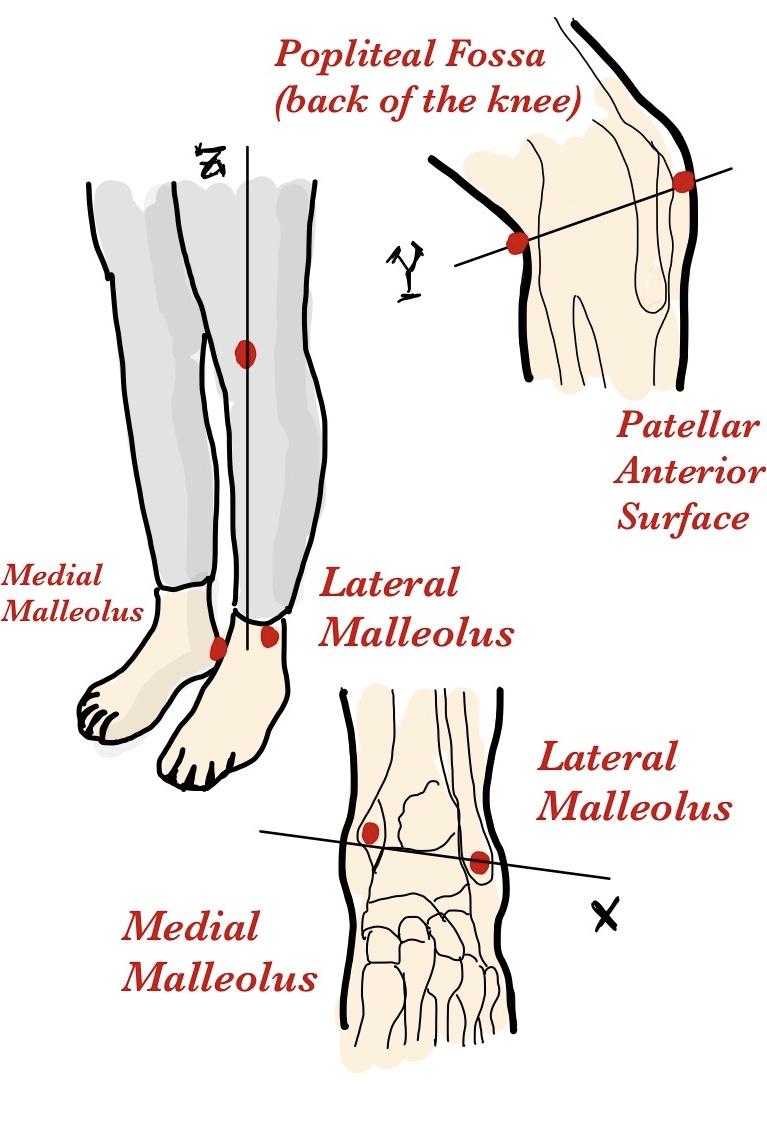} \\
\addlinespace[2pt]
14,15 & (R/L) Foot & CT, FMH, LM, MM & X: MM $\leftrightarrow$ LM; Y: CT $\to$ FMH; Z: CT $\to$ MM-LM line & Ankle to toes & \includegraphics[height=0.7in]{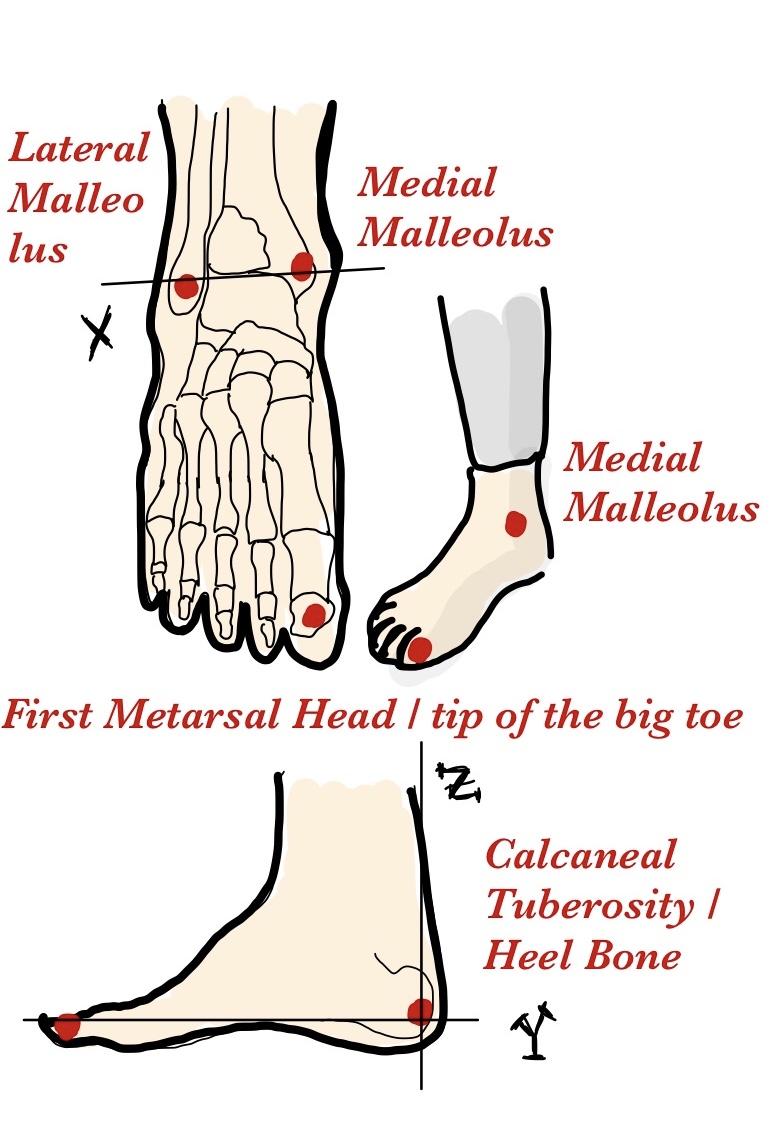} \\
\bottomrule
\multicolumn{6}{@{}p{\linewidth}@{}}{\vspace{0.3em}\textbf{Abbreviations:} LHJ/RHJ = Left/Right Helix-Tragus Junction; MMP = Midpoint between Mastoid Processes; LMP/RMP = Left/Right Mastoid Process; LAP/RAP = Left/Right Acromion Process; ASIS = Anterior Superior Iliac Spine; PSIS = Posterior Superior Iliac Spine; RSP = Radial Styloid Process; USP = Ulnar Styloid Process; MCP = Third Metacarpophalangeal Joint; MHE/LHE = Medial/Lateral Humerus Epicondyle; FME/FLE = Femur Medial/Lateral Epicondyle; LM/MM = Lateral/Medial Malleolus; CT = Calcaneal Tuberosity; FMH = First Metatarsal Head.} \\
\end{tabular*}
\end{table}
\end{landscape}

\end{document}